# Spectroscopic Signatures of Nonlocal Interfacial Coupling in Superconducting FeSe/SrTiO$_3$ Heterostructures

Nina Andrejevic[1†], Fei Han[2†], Thanh Nguyen[2], Alexander Puretzky[3], Qingping Meng[4], Yi-Fan Zhao[5], Weiwei Zhao[6], Lijun Wu[4], David Geohegan[3], Cui-Zu Chang[5], Yimei Zhu[4*], Shengxi Huang[7*], and Mingda Li[2*]

[1]Department of Materials Science and Engineering, Massachusetts Institute of Technology, Cambridge, MA 02139
[2]Department of Nuclear Science and Engineering, Massachusetts Institute of Technology, Cambridge, MA 02139
[3]Center for Nanophase Materials Sciences, Oak Ridge National Laboratory, Oak Ridge, TN 37831
[4]Department of Condensed Matter Physics and Materials Science, Brookhaven National Laboratory, Upton, NY 11973
[5]Department of Physics, The Pennsylvania State University, University Park, PA 16802
[6]Harbin Institute of Technology, Shenzhen, China
[7]Department of Electrical Engineering, The Pennsylvania State University, University Park, PA 16802
*mingda@mit.edu, *sjh5899@psu.edu, *zhu@bnl.gov
†N.A. and F.H. contributed equally to this work.

**Abstract** The mechanism of enhanced superconductivity in the one unit-cell (1UC) FeSe film on a SrTiO$_3$ (STO) substrate has stimulated significant research interest but remains elusive. Using low-temperature, voltage-gated Raman spectroscopy and low-temperature valence electron energy loss spectroscopy (VEELS), we characterize the phonon behavior and interfacial charge transfer in single- and few-layer FeSe films on STO. Raman measurements reveal ambipolar softening of the FeSe vibrational modes, mimicking that of the underlying STO substrate. We attribute this behavior to an interfacial coupling effect of STO on FeSe lattice dynamics. This interfacial coupling effect is further supported by local electron effective mass enhancement, which is determined from the red-shift in the FeSe VEELS spectrum near the FeSe/STO interface. Our work sheds light on the possible interfacial mechanisms contributing to the enhanced superconductivity across the FeSe/STO interface and further unveils the potential of low-temperature gated Raman spectroscopy and VEELS in clarifying a broad category of quantum materials.

The discovery of iron-based superconductivity [1] has attracted widespread attention due to its high superconducting transition temperature ($T_c$) [2–6] and novel electronic phase diagram in which spin density wave (SDW) and superconducting orders coexist and compete [7–9]. It is widely accepted that superconductivity in these materials originates from antiferromagnetic spin fluctuations in the so-called s± wave Cooper pairing state. This state is characterized by nodeless and nearly isotropic order parameters with opposite sign on the nested electron and hole Fermi surfaces [10–12]. That being said, the absence of hole pockets in intercalated or 1UC FeSe [13–20] challenges the former understanding that coexistence of electron and hole pockets is instrumental to superconductivity, which emerges after the suppression of the SDW instability. In particular, when 1UC FeSe is interfaced with a heat-treated SrTiO3 (STO) substrate, the $T_c$ is found to greatly exceed that of bulk FeSe, with a superconducting gap up to 20 meV [21] and is among the highest for superconducting heterostructure systems [22]. This has led to many experimental investigations of the interface effect [18, 21, 23–25], including the roles of interfacial charge transfer and electron-phonon coupling. However, the detailed mechanism at a mode- and spatially-resolved level is not yet clear.

In this work, we study the phonon dynamics and interfacial charge transfer in 1UC FeSe/STO using low-temperature, voltage-gated Raman spectroscopy and low-temperature valence electron energy loss spectroscopy (VEELS), respectively. The gate voltage modulates the density of proximity charge carriers at the FeSe/STO interface, while Raman spectroscopy provides an extremely sensitive tool to study the interfacial electron-phonon coupling phenomena at a mode-resolved level. On the other hand, low-temperature VEELS offers a unique tool to directly quantify the charge transfer process with atomic resolution. Ambipolar softening of FeSe phonons with gate voltage is observed, mirroring that of the underlying substrate. In addition, an unusual red-shift in the FeSe plasmon peak is discernable near the FeSe/STO interface. Both suggest an unusual nonlocal interplay between the FeSe charge carriers and STO phonons. Our work helps to elucidate the interfacial mechanism in the superconducting FeSe/STO heterostructures and demonstrates a promising pathway to utilize low-temperature gated Raman spectroscopy and VEELS to study electron-phonon interaction mechanisms in a broad category of quantum materials.

Raman spectroscopic measurements were carried out on superconducting FeTe-capped 1UC FeSe films atop TiO$_2$-terminated insulating STO (001) substrates. Comparison between the Raman spectra of FeTe/1UC FeSe/STO and bare STO at room temperature reveals additional peaks contributed by FeSe and FeTe vibrational modes in the range of 100 - 200 cm$^{-1}$, as shown in **Figure 1a**. Most prominent are the peaks found at 104, 126, and 143 cm$^{-1}$ in the 300K spectrum, which we attribute to the FeSe E$_g$, FeTe A$_{1g}$, and FeTe B$_{1g}$ modes, respectively, shown in **Figures 1b** and **c**. Strong signatures of these additional peaks likewise appear in the temperature-dependent Raman spectra of the few-layer sample consisting of an 8UC FeSe thin film (**Figure S1b**), corroborating the assignment. Superconductivity in both films was confirmed by transport measurements (**Figure S5**). The two sharp Raman peaks at 126 and 143 cm$^{-1}$ identify with the FeTe A$_{1g}$ and B$_{1g}$ modes [26, 27], while the comparatively weak peak at 104 cm$^{-1}$ is consistent with the FeSe $E_g$ mode found in related work [28] and with the thin layer of FeSe present. We also note subtle evidence of the FeSe $A_{1g}$ and $B_{1g}$ modes near 166 and 192 cm$^{-1}$ which are likewise in good agreement with existing works [29, 30]. The Raman signatures of these vibrational modes are apparent even at low temperatures down to 4K (**Figure 1a**) and are distinguishable from the STO Raman peaks, which differ significantly between 300K and 4K due to the well-known structural transition near 110K [31].

Notably, the positions of the FeSe and FeTe Raman peaks in this energy range are quite insensitive to temperature when compared with the behavior in bulk crystals [28, 29], which underscores the significance of the STO substrate in templating the growth of the FeSe layer. Such structural stabilization of FeSe by the substrate was also shown to be favorable for enhanced electron-phonon coupling in this system [32]. Having identified the relevant vibrational modes, we then varied a backgate voltage $V_g$ between -200V and 200V across the FeTe/1UC FeSe/STO sample, which was maintained at 4K, well below $T_c$ [21, 33] (schematics in **Figure 1d**). Qualitative changes in the spectral intensity as a function of $V_g$ are highlighted in **Figure 1e** and shown to be highly reproducible.

With all vibrational modes assigned, we plot all five measured spectra from negative to positive bias voltage in **Figure 2a**. To characterize the phonon behavior as a function of gate voltage, we directly determined the peak positions from the raw Raman spectra by identifying zeros and quasi-discontinuities in its first derivative, which correspond to peaks and kinks, respectively, as shown in **Figures 2b** and **c** for FeSe and FeTe vibrational modes, and **Figures S2a** and **b** for STO modes. We first discuss the gate dependence of FeSe and FeTe Raman shifts (**Figure 2d**) with corresponding ionic vibrations (**Figure 2e**). The FeSe $E_g$ mode corresponds to Se vibrations in the plane perpendicular to the direction of the applied electric field, while the FeSe and FeTe $A_{1g}$ and $B_{1g}$ modes comprise vibrations of either the chalcogen or metal atom, respectively, parallel to the electric field direction. We make this distinction in order to separate the effect of the static electric field on material polarization [34] from that of electrostatic doping on electron-phonon coupling, both of which may be reflected in Raman spectra as changes in the peak frequency. Despite structural similarities between FeSe and FeTe, the latter is largely insensitive to the gate voltage except for the intensity reduction of the $A_{1g}$ mode (**Figure 2g**). By contrast, FeSe vibrational modes exhibit systematic ambipolar softening with applied gate voltage. This is especially apparent in the $A_{1g}$ and $B_{1g}$ Raman shifts, whose intensities also show slight increase with gate voltage. The corresponding behavior of STO modes is shown in **Figure S2c**, where we remark the emergence of the "silent" STO mode near 261 cm$^{-1}$, which is only Raman-active in the presence of external electric fields [35], as an indicator of our applied gate voltage. We see that the 400 - 500 cm$^{-1}$ window, which hosts the STO $E_g$+$B_{1g}$ and LO$_2$ modes [36], similarly exhibits ambipolar softening and increasing intensity (**Figures 2f**, **S2**, and **S3**). The remaining STO modes are either slightly hardened in either direction or left unchanged.

The ambipolar dependence of STO modes on gate voltage is highly reminiscent of the electric field-induced strain dependence in STO [37], which is characteristic of a piezoelectric at voltages near ±200V. Thus, the Raman response of STO under the backgate potential is likely due to strain induced by the inverse piezoelectric effect. Since the applied bias is expected to fall primarily over the STO substrate in the backgate configuration, we suggest that ambipolar softening mirrored by the adjacent FeSe modes is caused in large part by interlayer coupling to STO. While the effect of the STO substrate on FeSe has been explored in the context of (FeSe) electron-(STO) phonon coupling [30, 38], our results further show that coupling to STO can also affect the lattice dynamics of FeSe. Interestingly, we also observe slightly more softening of the FeSe $B_{1g}$ mode for positive than negative bias (**Figure 2d**). This may indicate a strengthened electron-phonon interaction akin to that proposed in recent work [23, 33], which showed that a positive backgate was favorable for attracting transferred electrons in the FeSe film to the FeSe/STO interface and thereby enhanced their coupling to interfacial phonons without a significant change to the carrier density. Thus, the intimate connection between FeSe and the STO interface promotes the intriguing behavior of FeSe phonons in this system.

To corroborate the observations from Raman measurements, we performed low-temperature VEELS measurements on both 1UC and 8UC FeSe samples at T=10K to image the spatially-dependent charge transfer and redistribution process. The spectra were taken along a line across the FeSe/STO interfaces shown in **Figures 3b** and **c**. Considering the FeSe/STO interface as a prototypical metal/semiconductor junction, the contact potential difference is expected to develop a depletion layer in STO near the interface and a complementary, local accumulation of electrons in FeSe. This electron transfer from STO to FeSe has also been suggested experimentally from a signature blue-shift of the Fe $L_3$ edge in the core-level EELS [33]. By contrast, the VEELS data reveal a red-shift in both the STO and FeSe plasmon peaks as one approaches the FeSe/STO interface, for both 1UC and 8UC FeSe samples, as indicated in **Figures 3d-f** and in **Figures S4a-d**.

Our observation contradicts the expected behavior of the FeSe plasmon peak obtained from dielectric function theory calculations, which instead predict a slight blue-shift, as shown in **Figures 3e-f**. The red-shifts on either side of the interface could be attributed either to local depletion in electron concentration or to local enhancement of the electron effective mass. If we suppose that the red-shift in the STO plasmon peak arises from the transfer of interfacial electrons to FeSe, then it is possible that the red-shift of the FeSe plasmon peak is contributed at least in part by a local enhancement of the FeSe electron effective mass mediated by interaction with STO phonons at the interface. However, additional experiments would be necessary to further clarify this point. The VEELS measurement clearly reveals an interfacial effect on FeSe electrons within a ∼ 1nm-wide region, comparable to the width of the first two atomic FeSe layers which are believed to host superconductivity from thickness-dependent studies [18, 39–41].

To summarize, we used complementary low-temperature gated Raman spectroscopy and VEELS measurements to characterize the FeSe phonon dynamics and interfacial charge transfer in FeTe-capped 1UC and 8UC FeSe films on STO. Despite the structural similarities of FeSe and

FeTe, we identify a unique ambipolar response of the FeSe vibrational modes that mimics the underlying substrate and indicates a nonlocal interfacial coupling effect of STO on FeSe lattice dynamics. The importance of interfacial coupling is strengthened by our observation of an unexpected red-shift in the FeSe VEELS spectrum near the FeSe/STO interface, which is a possible signature of FeSe electron effective mass enhancement through interaction with STO interfacial phonons. Our work adds new insights to the rich and intriguing study of high-Tc superconductivity in the 1UC FeSe/STO system and demonstrates the application of low-temperature gated Raman spectroscopy and VEELS to probe the electron-phonon characteristics of a quantum material.

## Methods

*Sample Growth and Preparation* The FeSe films used in this work were grown in an ultrahigh vacuum (UHV) MBE chamber with a base pressure lower than $5\times10^{-10}$ mbar. The insulating STO(001) substrates were heat-treated to obtain a uniform $TiO_x$-terminated surface. Then, the STO substrates were transferred into the MBE chamber and degassed at 600°C for 1h. FeSe films were grown by co-evaporating Fe(99.995%) and Se(99.999%) from two separate Knudsen cells with a flux ratio of 1:20 on the STO substrate at 330°C. The growth rate for the films was approximately 0.25 UC/min. Epitaxial growth was monitored by *in situ* reflective high-energy electron diffraction (RHEED), where the high crystal quality and the atomically flat surface were confirmed by the streaky and sharp "1×1" patterns. To avoid possible contamination, a 10UC-thick epitaxial FeTe capping layer and another 10nm Te layer were deposited on top of the FeSe films before removal from the growth chamber for gate Raman spectroscopy and VEELS measurements.

*Gated Raman Spectroscopy* The Raman spectra were measured in a custom-built micro-Raman setup. The samples were excited with a continuous wave (CW) diode-pumped solid-state laser (Excelsior, Spectra Physics, 532nm, 100 mW) through an upright microscope using a 50x long-working distance objective with NA (numeric aperture) = 0.5. The typical incident laser power on a sample was maintained at ~ 100 μW to reduce possible laser heating and damaging of the samples during Raman spectra acquisition. The scattered Raman light was analyzed by a spectrometer (Spectra Pro 2300i, Acton, f = 0.3 m) that was coupled to the microscope and equipped with an 1800 grooves/mm grating and a CCD camera (Pixis 256BR, Princeton Instruments). The low-temperature Raman spectra were measured using a liquid He-cryostat (MicrostatHiResII, Oxford Instruments) with a temperature controller (MercuryiT, Oxford Instruments) that allowed precise temperature control from 3.6 to 300 K as well as biasing a sample. The cryostat was mounted on a motorized XY microscope stage (Marzhauser) under the microscope of the micro-Raman setup. The cryostat was evacuated to the base pressure of $7\cdot10^{-7}$ mbar prior to cool down.

*Valence Electron Energy Loss Spectroscopy* The VEELS spectra are analyzed to reveal the electronic state of the interfacial layers. Each spectrum was deconvoluted using the Fourier-log method to remove the zero-loss peak and the effect of plural scattering [42]. Complex dielectric constants of STO, FeSe, and FeTe in position coordinates can be extracted from a Kramers-Kronig analysis using experimentally-measured EELS. The formulas used in our calculation follow Appendix II of Moreau *et al.* [43]. Although Moreau *et al.* include interfacial effects based on dielectric theory, they assume that the interface is an ideal geometrical plane and that the dielectric constant behaves as a step function of position across the interface of the two media. However, the actual intermediate layer cannot be sharp [44] due to the likely occurrence of charge transfer upon contact of the two media. The EELS of the low-loss region (up to 50 eV) is dominated by plasmons. The plasmon peaks in VEELS are determined by fitting the data to a superposition of Lorentzian functions. The plasmon peak position is given by [45]:

$$\omega_p^2 = \frac{4\pi e^2 N_{eff}}{m(1+\delta\varepsilon_0)} \qquad (1)$$

where

$$N_{eff} = \frac{m}{m^*}N_v + \sum_{l>v} f_{lv}, \qquad (2)$$

and $e$ is the elementary charge; $m$ and $m^*$ are respectively the mass and the effective mass of electron; and $N_v$ is the number of valence electrons which are essentially unbound in the frequency range. The second term $\sum_{l>v} f_{lv}$ in $N_{eff}$ is due to the valance-band electrons from the interband transition. $\delta\varepsilon_0$ describes the real interband transitions and is zero until the real excitations from the bound band are allowed [45, 46]. Consequently, we ignore the effect of $\delta\varepsilon_0$ in our calculations. From Eq. 1, the variation of the plasma frequency caused by the change in electron density can be determined as:

$$\Delta N = \frac{m}{4\pi e^2}\left[\omega_{pe}^2(x) - \omega_{pc}^2(x)\right] \qquad (3)$$

where $\omega_{pe}$ and $\omega_{pc}$ are plasmon frequencies in coordinate $x$ from experimental and calculated VEELS spectra, respectively.

*Transport Measurement* The Hall and longitudinal resistances were carried out in a Quantum Design Physical Property Measurement System (1.8 K, 9 T) with the excitation current flowing in the film plane and the magnetic field applied perpendicular to the plane. The FeSe/STO films were scratched to a six-terminal Hall bar geometry device using a needle by hands. The backgate voltage was applied using the Keithley 2450.

## Acknowledgements

N.A. acknowledges the support of the National Science Foundation Graduate Research Fellowship Program under Grant No. 1122374. T.N. thanks the support from the MIT SMA-2 Fellowship Program. C.Z.C. acknowledges the support from NSF-CAREER award (DMR-1847811). Raman spectroscopic measurements were conducted at the Center for Nanophase Materials Sciences, which is a DOE Office of Science User Facility. Electron microscopic and spectroscopic measurements used resources of the Center for Functional Nanomaterials, which is a U.S. DOE Office of Science Facility, at Brookhaven National Laboratory under Contract No. DE-SC0012704.

## Conflict of Interest

The authors declare no conflict of interest.

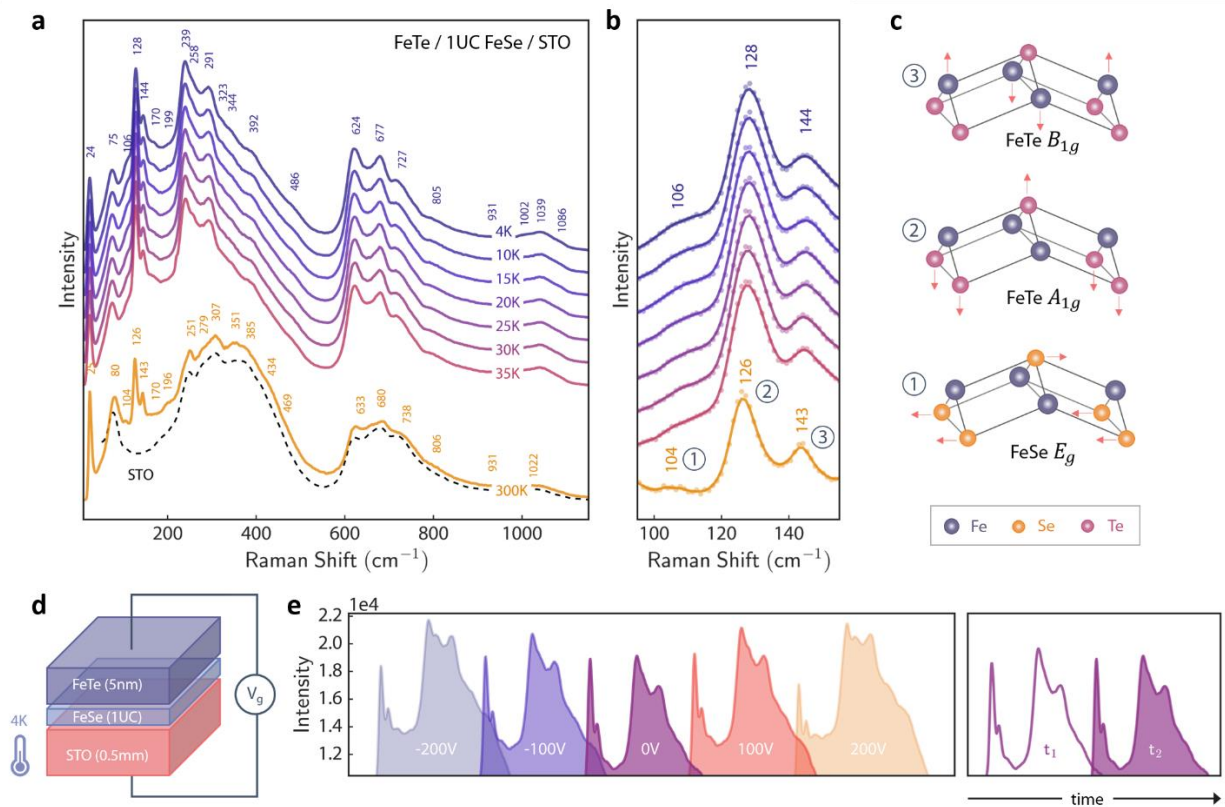

**Figure 1. Temperature-dependent Raman spectroscopic measurements. a.** Temperature-dependent Raman spectra of FeTe/1UC FeSe/STO from 4K to 35K and at room temperature. Spectra have been vertically offset for clarity. Numbers mark various peaks of the 4K and 300K spectra, and the black dashed line gives the bare STO spectrum at 300K. **b.** Enlarged view of the Raman spectra in **a** at the peak locations of the FeSe $E_g$ and FeTe $A_{1g}$ and $B_{1g}$ vibrational modes. **c.** Ionic vibrations of the FeSe $E_g$, FeTe $A_{1g}$, and FeTe $B_{1g}$ modes. The numbers correspond to the peaks in **b**. **d.** Schematic illustration of the FeTe/1UC FeSe/STO sample, consisting of a thick (0.5 mm) STO substrate and ∼5nm FeTe capping layer. Gate voltage-dependent Raman spectra were collected at a fixed temperature of T=4K. **e.** The left panel indicates the qualitative intensity differences among spectra collected under different gate voltages from -200V to 200V. Spectra are shown in the range of 100 - 500 cm$^{-1}$ and have been horizontally offset for clarity. The right panel indicates the reproducibility of spectra taken under the same conditions, but at different time separated by several measurements at various gate voltages.

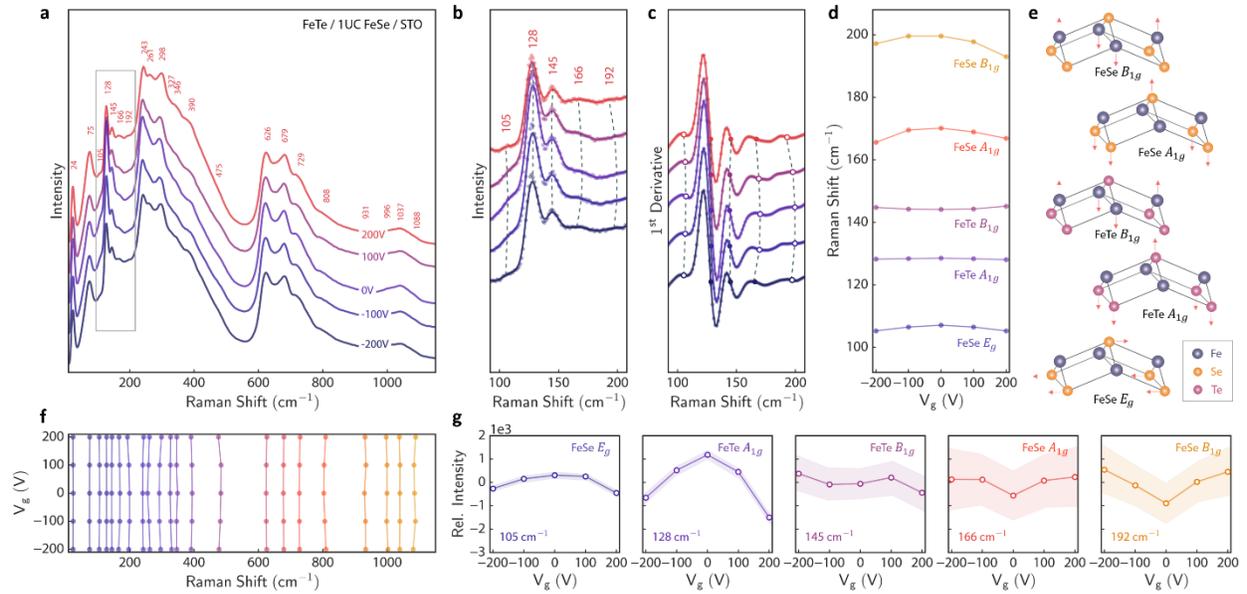

**Figure 2. Gated Raman spectroscopy at low temperature. a.** Voltage-dependent Raman spectra of FeTe/1UC FeSe/STO at a fixed temperature of 4K. Spectra have been vertically offset for clarity, and numbers mark various peaks of the 200V spectrum. **b.** Enlarged view of the Raman spectra in a within the boxed region, highlighting the Raman peaks from FeSe and FeTe vibrational modes. Changes in peak positions with gate voltage are traced by the dashed lines. **c.** First derivative of the spectra in **b** used to identify the locations of peaks (filled circles) and kinks (open circles) in the original spectra as a function of gate voltage. **d.** Raman peak positions of the modes in b as a function of gate voltage, indicating the ambipolar softening of the FeSe modes contrasted against the relatively constant FeTe peak positions. **e.** Ionic vibrations of the FeSe $E_g$, $A_{1g}$, and $B_{1g}$ vibrational modes, and FeTe $A_{1g}$ and $B_{1g}$ vibrational modes portrayed in **b-d**. **f.** Raman peak positions of the full spectra in a as a function of the gate voltage. **g.** Raman peak intensity of the modes in **b** as a function of the applied gate voltage. Numbers in the lower-left corner indicate the mode frequency at 200V bias.

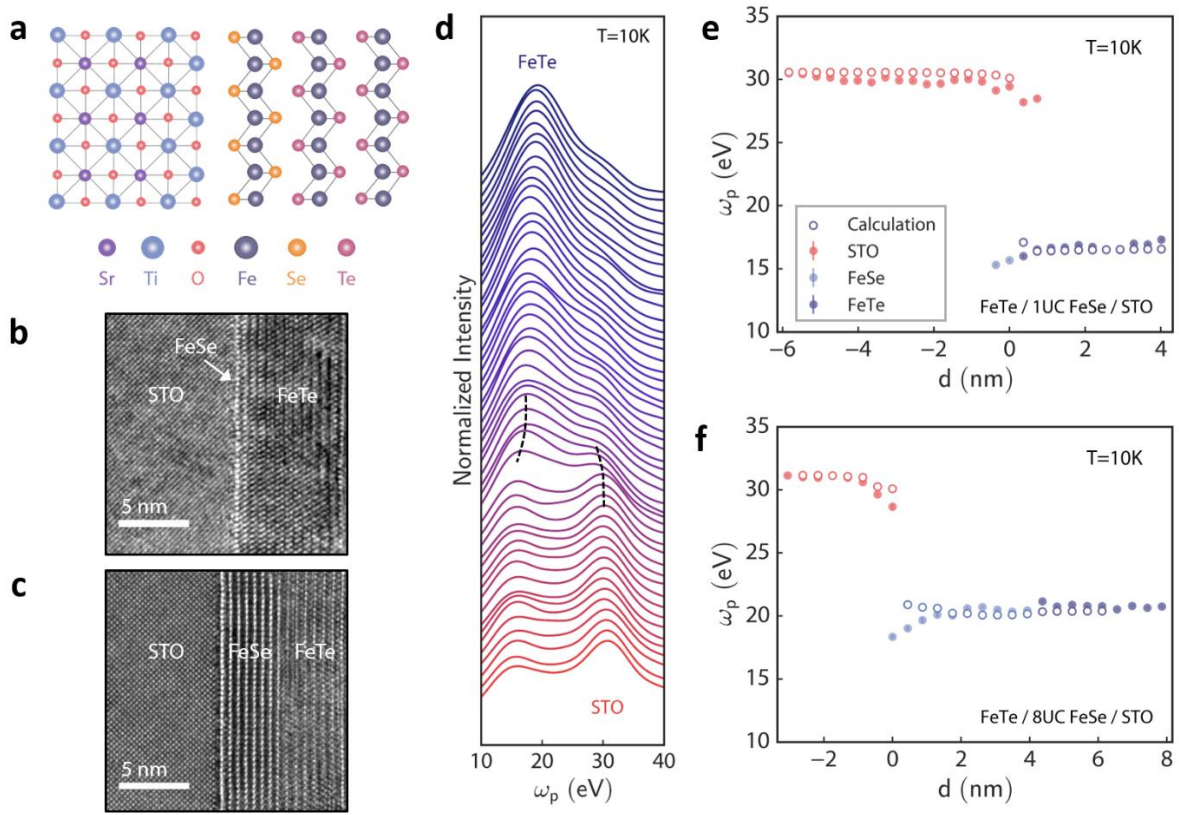

**Figure 3. Valence electron energy loss spectroscopy. a.** Schematic illustration of the FeTe/1UC FeSe/STO atomic structure. High-resolution transmission electron micrographs of **b.** FeTe/1UC FeSe/STO and **c.** FeTe/8UC FeSe/STO. **d.** Series of VEELS spectra scanned perpendicularly across the interfaces from the STO bulk to the FeTe bulk. The total scan length is ∼ 10nm. Black dashed lines indicate a red-shift in both the STO and FeSe plasmon peaks near the FeSe/STO interface. **e.** Plasmon frequency as a function of the position coordinate along a path perpendicular to the sample interfaces in FeTe/1UC FeSe/STO. Experimental and calculated values are plotted as filled and open circles, respectively, for STO (red), FeSe (light blue), and FeTe (indigo). The corresponding plot for FeTe/8UC FeSe/STO is given in **f.**

## Supplementary Figures

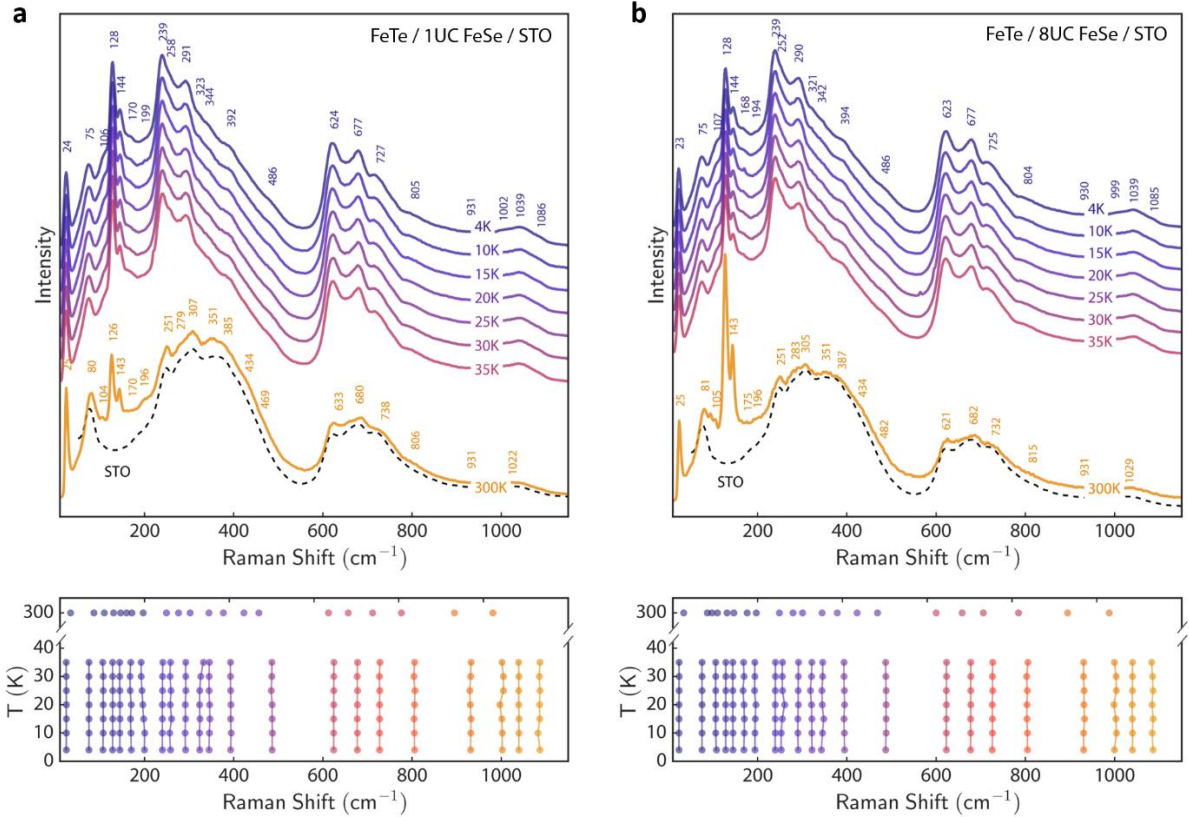

**Figure S1. Temperature-dependent Raman spectroscopy of single- and few-layer FeSe thin films.** Temperature-dependent Raman spectra of **a.** FeTe/1UC FeSe/STO and **b.** FeTe/8UC FeSe/STO from 4K to 35K and at room temperature. Spectra have been vertically offset for clarity. Numbers mark various peaks of the 4K and 300K spectra, and the black dashed line gives the bare STO spectrum at 300K. Lower panels indicate the temperature evolution of the Raman peak positions.

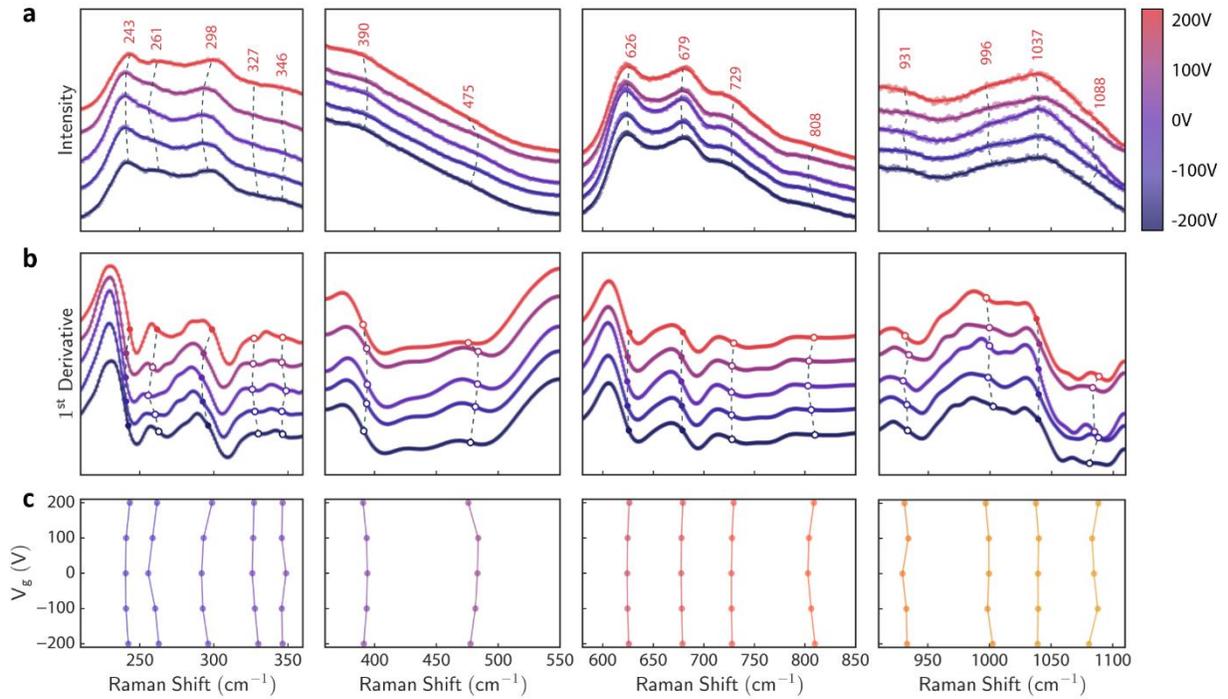

**Figure S2. Gate-dependent STO Raman peak positions. a.** Enlarged views of voltage-dependent Raman spectra of FeTe/1UC FeSe/STO at a fixed temperature of 4K within four different windows (4 panels), each highlighting Raman peaks from STO vibrational modes. Spectra have been vertically offset for clarity, and numbers mark various peaks of the 200V spectrum. Changes in peak positions with gate voltage are traced by the dashed lines. **b.** First derivative of the spectra in a used to identify the locations of peaks (filled circles) and kinks (open circles) in the original spectra as a function of gate voltage. **c.** Raman peak positions of the modes in b as a function of gate voltage.

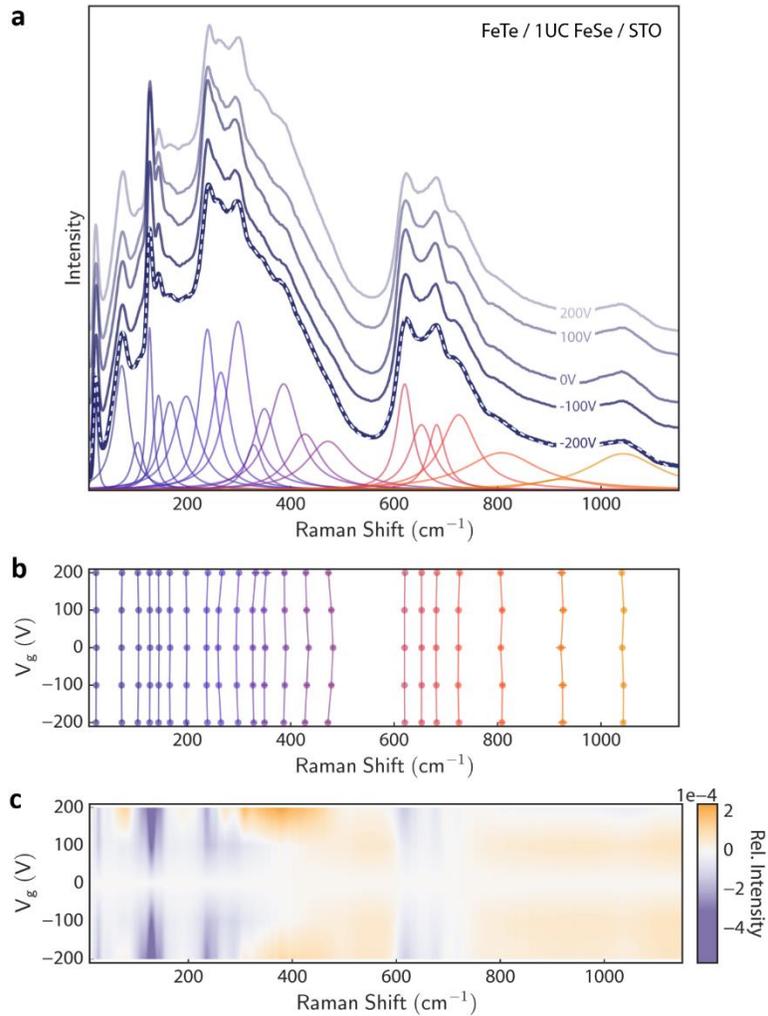

**Figure S3. Peak fitting of gated Raman spectra. a.** Voltage-dependent Raman spectra of FeTe/1UC FeSe/STO at a fixed temperature of 4K are reproduced from **Figure 2a** with a vertical offset for clarity. The dashed white line through the bottom-most spectrum is a representative peak-fit of a single Raman spectrum (Vg = -200V) by a sum of Lorentzian functions. Individual Lorentzian line shapes constituting the fit are plotted in a color gradient along the bottom. **b.** Raman peak positions obtained by fitting the spectra in a are plotted as a function of the gate voltage. **c.** Intensity map of the area-normalized Raman spectra in a as a function of the Raman shift and gate voltage. The map is in qualitative agreement with individual peak intensities obtained from peak fitting, which are depicted in **Figure 2g**.

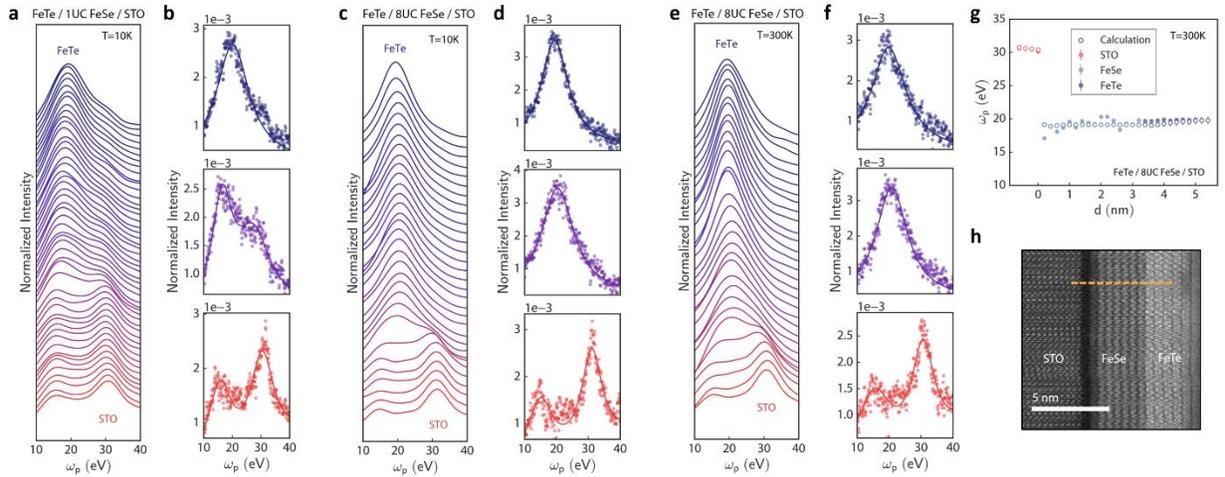

**Figure S4. Valence electron energy loss spectroscopy.** Series of VEELS spectra scanned perpendicularly across the interfaces from the STO bulk to the FeTe bulk for **a.** FeTe/1UC FeSe/STO at 10K, **c.** FeTe/8UC FeSe/STO at 10K, and **e.** FeTe/8UC FeSe/STO at 300K. To the right of the corresponding waterfall plot, vertical tri-panel plots in **b**, **d**, and **f** show representative VEELS spectra taken at 3 individual points along the line scan jointly with the peak fit used in plotting the smooth curves shown in **a**, **c**, and **e**. **g.** Plasmon frequency as a function of the position coordinate along a path perpendicular to the sample interfaces in FeTe/8UC FeSe/STO at 300K. Experimental and calculated values are plotted as filled and open circles, respectively, for STO (red), FeSe (light blue), and FeTe (indigo). **h.** High-resolution transmission electron micrograph of FeTe/8UC FeSe/STO at 300K. Dashed orange line corresponds to path along which the scans shown in **e** and **g** were made.

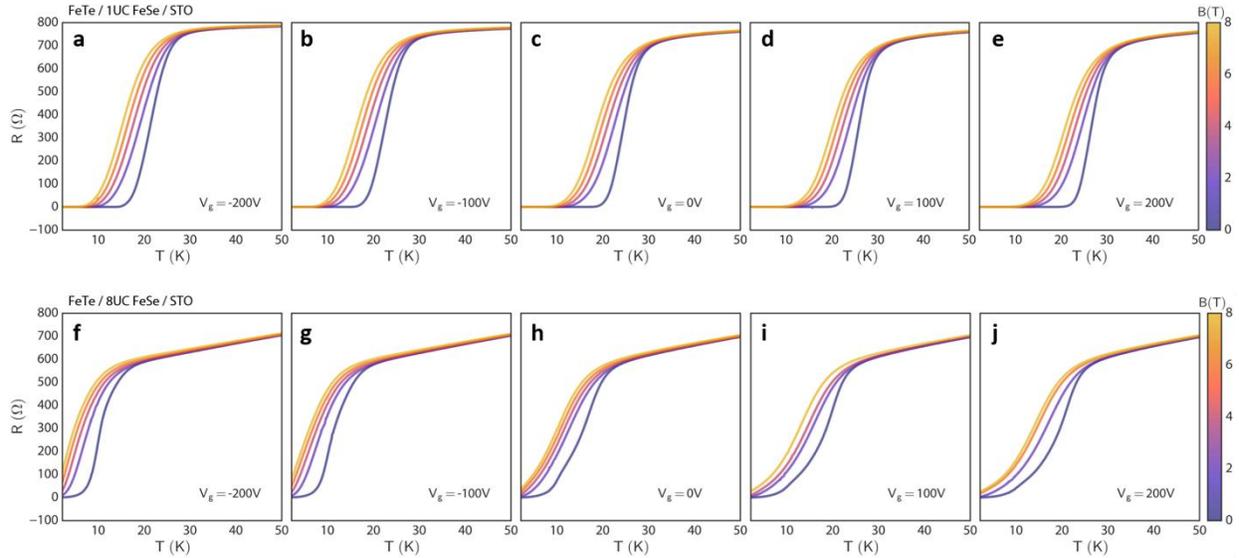

**Figure S5. Transport characteristics under varied magnetic field at different gate voltages.** The evolution of temperature-dependent resistance in 1UC (**a-e**) and 8UC (**f-j**) FeSe films on STO with magnetic field is shown for several gate voltages between -200V and 200V, confirming superconductivity is attained at all relevant experimental conditions in both samples. Figures are reproduced with data from [S1].

[S1] W. Zhao et al., Science Advances 4 (2018), 10.1126/sciadv.aao2682.